\documentclass[iop]{emulateapj}

\newcommand{\hMsun}{{\ifmmode{h^{-1}{\rm {M_{\odot}}}}\else{$h^{-1}{\rm{M_{\odot}}}$~}\fi}}
\newcommand{\hMpc}{{\ifmmode{h^{-1}{\rm Mpc}}\else{$h^{-1}$Mpc }\fi}}

\shorttitle{Redshift Dependent AP Test: Galaxy Density Gradient Field}
\shortauthors{X.-D. Li, C. Park, J.E. Forero-Romero, J. Kim}

\begin{document}


\title{Cosmological constraints from the redshift dependence of the Alcock-Paczynski test: galaxy density gradient field}



\author{Xiao-Dong Li\altaffilmark{1}, Changbom Park\altaffilmark{2},}
\affil{School of Physics, Korea Institute for Advanced Study, 85 Heogiro, Dongdaemun-gu, Seoul 130-722, Korea}
\author{J.~E.~Forero-Romero\altaffilmark{3},}
\affil{Departamento de F\'{i}sica, Universidad de los Andes, Cra. 1 No. 18A-10, Edificio Ip, Bogot\'a, Colombia}
\and
\author{Juhan Kim\altaffilmark{4}}
\affil{Center for Advanced Computation, Korea Institute for Advanced Study, 85 Hoegi-ro, Dongdaemun-gu, Seoul 130-722, Korea}

\altaffiltext{1}{xiaodongli@kias.re.kr}
\altaffiltext{2}{cbp@kias.re.kr}
\altaffiltext{3}{je.forero@uniandes.edu.co}
\altaffiltext{4}{kjhan@kias.re.kr}


%
%
%
%
%


\begin{abstract}
We propose a method based on the redshift dependence of the
Alcock-Paczynski (AP) test to measure the expansion history of the
Universe. 
It uses the isotropy of the galaxy density gradient field to constrain
cosmological parameters. 
If the density parameter $\Omega_m$ or the dark energy equation of
state $w$ are incorrectly chosen, 
the gradient field appears to be anisotropic 
with the degree of anisotropy varying with redshift. 
We use this effect to constrain the cosmological parameters governing
the expansion history of the Universe. 
Although redshift-space distortions (RSD) induced by galaxy peculiar
velocities also produce anisotropies in the gradient field, these
effects are close to uniform in magnitude over a large
range of redshift.
This makes the redshift variation of the gradient field anisotropy 
relatively insensitive to the RSD.    
By testing the method on mock surveys drawn 
from the Horizon Run 3 cosmological N-body simulations,
we demonstrate that the cosmological parameters can be estimated without bias.
Our method is complementary to the baryon acoustic oscillation or
topology methods as it depends on $D_AH$, the product of the angular
diameter distance and the Hubble parameter. 
\end{abstract}





\section{Introduction}

Since the discovery of cosmic acceleration \citep{Riess1998,Perl1999},
the idea of cosmological constant or dark energy has been used in cosmology
as a theoretical explanation of the phenomenon.
To constrain theories further it is crucial to increase the amount of observational
data and to improve the statistical methods for measuring the cosmological parameters governing the expansion of the Universe.

The Alcock-Paczynski (AP) test \citep{AP1979} is a pure geometric probe of the cosmic expansion history
based on comparison of observed tangential and radial dimensions of objects
which are known to be isotropic.
There are a few methods proposed to apply the AP test for cosmological purposes.
The most widely adopted one is the method using anisotropic clustering \citep{Ballinger1996,Matsubara1996},
which has been used for the 2-degree Field Quasar Survey \citep{Outram2004},
the WiggleZ dark energy survey \citep{Blake2011}, the SDSS-II LRG survey \citep{ChuangWang2012},
and the SDSS-III Baryon Oscillation Spectroscopic Survey (BOSS) \citep{Reid2012,Corredoira2013,Anderson2013,Beutler2013,Chuang2013,Sanchez2013,Linder2013,Samushia2014}.
The main caveat of this method is that,
because the radial distances of galaxies are inferred from redshifts,
AP tests are inevitably limited by redshift-space-distortions (RSD) \citep{Ballinger1996},
which leads to apparent anisotropy even if the adopted cosmology is correct.
This effect must be then accurately modeled for the 2-point statistics of galaxy clustering.

A second interesting approach measures the symmetry properties of galaxy pairs \citep{Marinoni2010}.
Unfortunately this method is also seriously limited by peculiar velocities.
The RSD effect in the apparent tilt angles of galaxy pairs is found to be dependent on both redshift and underlying cosmology,
making it difficult to model accurately \citep{Jennings2011}.

\cite{Ryden1995} and \cite{LavausWandelt1995} proposed another method using the apparent stretching of voids.
It has the advantage that the void regions are easier to model compared with high density regions.
But this method also has limitations in that it utilizes only low density regions of the large scale structure
and requires much larger samples compared to other methods.

In this paper we propose a new method that overcomes these limitations.
It uses the distortion in the apparent density gradient field constructed from galaxy distribution.
The density gradient vectors are expected to be isotropic if the correct cosmology is adopted,
while an anisotropic distribution implies a wrongly assumed cosmology.
Similar to the 2-point statistics, this method is also based on the distribution of galaxies.
But our method samples the density gradient vectors uniformly within the survey volume,
and thus the high and low density regions are equally utilized.
In contrast, methods of 2-point statistics and galaxy pairs assign more weights to the high density regions,
while the void method utilizes low density regions only.

The observed density gradient field is also affected by RSD,
which perturbs galaxy positions along the line of sight (LOS) and produce spurious gradients.
However, we find that these anisotropies can be distinguished from those induced by wrongly assumed cosmological parameters
by looking at the redshift dependence of the anisotropies.
We conduct a proof-of-concept test of our method on the Horizon Run 3 (HR3) mock surveys.



This paper is organized as follows.
Section 2 briefly introduces the idea of the Alcock-Paczynski test.
Section 3 introduces the simulation data used in our analysis.
In section 4 and 5 the method is tested on the HR3 mock surveys.
We summarize and conclude in Section 6.

\begin{figure*}
   \centering{
   \includegraphics[height=8cm]{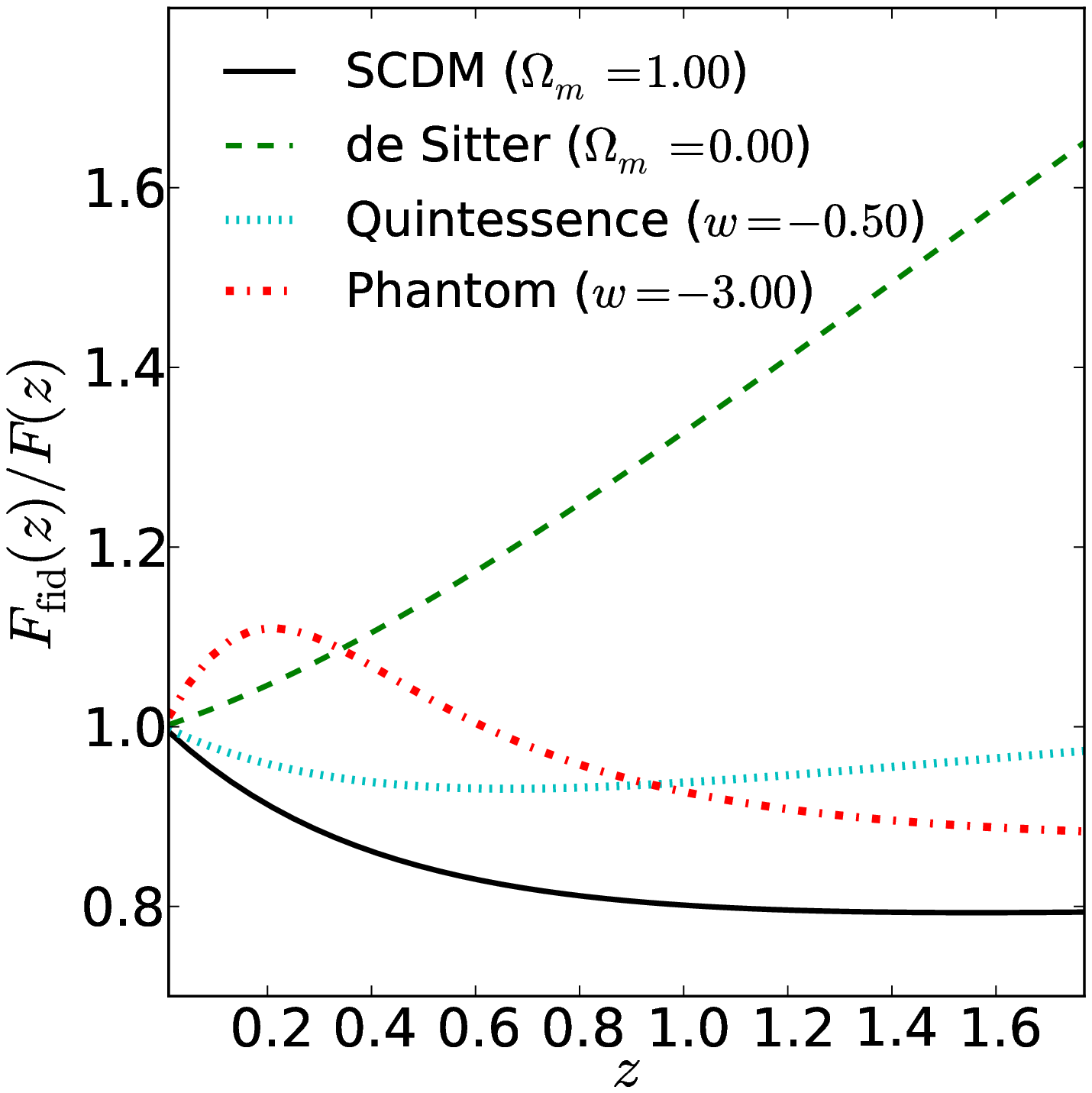}
   \includegraphics[height=8cm]{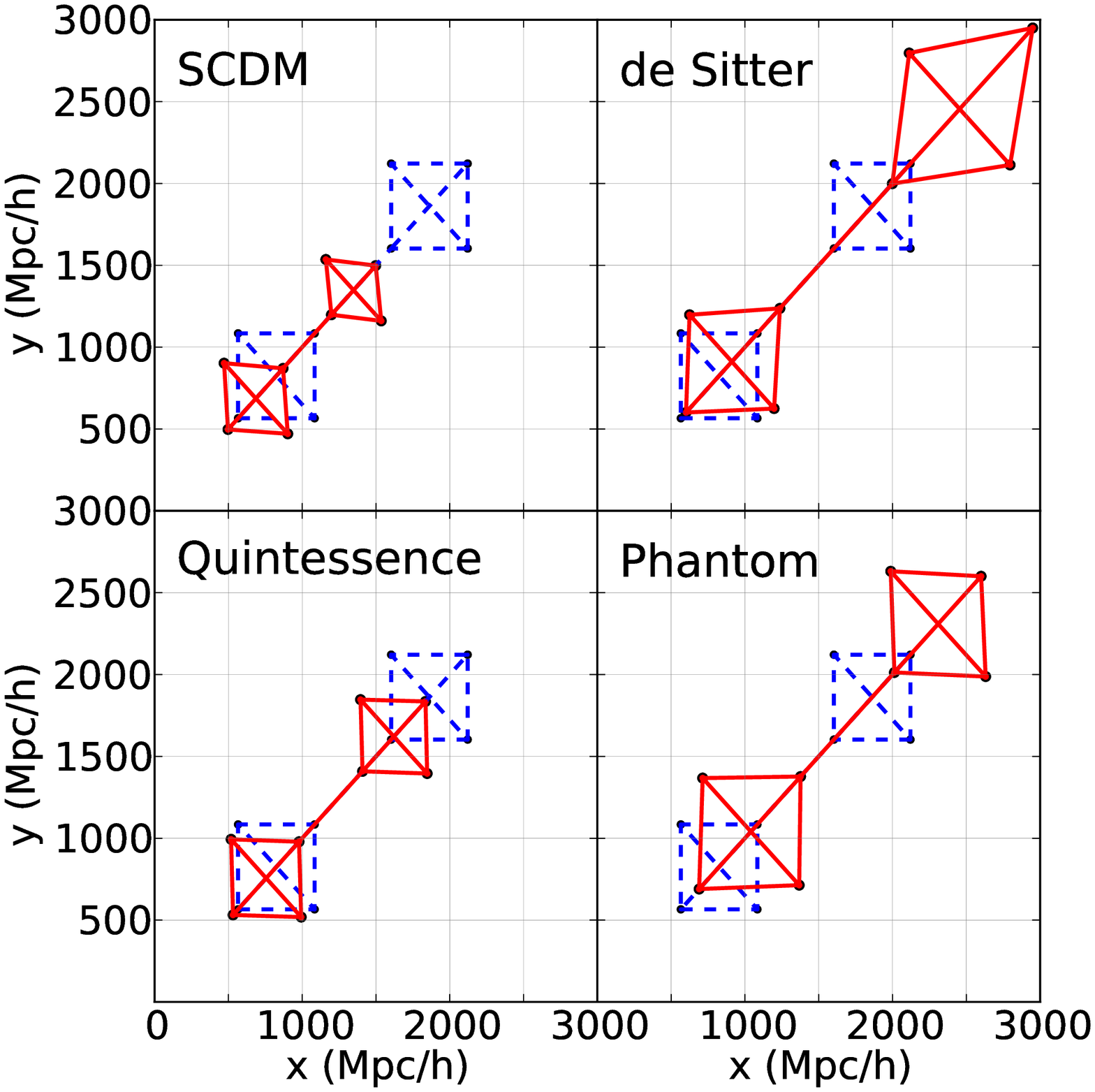}}
   \caption{\label{fig_AP}
   Apparent distortion of objects in four wrongly assumed cosmologies, assuming a true cosmology of $\Omega_m=0.26$, $w=-1$.
   {\it Left Panel:} Evolutions of the AP distortion parameter $F(z)\equiv(1+z)D_A(z)H(z)/c$,
   normalized by its value in the true cosmology.
   {\it Right Panel:} Apparent distortion of two perfect squares (red solid),
   measured by an observer located at the origin.
   For reference, blue dashed squares show their shapes and positions in the true cosmology.}
\end{figure*}

\section{The Alcock-Paczynski Test in a Nutshell}

Let us consider an object in the Universe for which the ratio between its sizes along and across the line-of-sight (LOS) is known.
In a given cosmology its observed redshift span $\Delta z$ and angular size $\Delta \theta$ are related with the comoving sizes by
\begin{equation}\label{eq:distance}
\Delta r_{\parallel} = \frac{c}{H(z)}\Delta z,\ \ \Delta r_{\bot}=(1+z)D_A(z)\Delta \theta,
\end{equation}
respectively, where $H$ is the Hubble parameter, $D_A$ is the angular diameter distance.
For simplicity let us consider a flat Universe composed of
a matter component with the present density parameter $\Omega_m$ and a dark energy component with constant equation of state (EoS) $w$.
We then have
\begin{eqnarray}\label{eq:HDA}
& &H(z) = H_0\sqrt{\Omega_ma^{-3}+(1-\Omega_m)a^{-3(1+w)}},\nonumber\\
& &D_A(z) = \frac{1}{1+z}r(z)=\frac{1}{1+z}\int_0^z \frac{c dz^\prime}{H(z^\prime)},
\end{eqnarray}
where $a=1/(1+z)$ is the cosmic scale factor,
$H_0$ is the present value of Hubble parameter and $r(z)$ is the comoving distance.
If we adopt correct values of cosmological parameters,
the inferred radial and tangential sizes of the object would be equal to
\begin{equation}\label{eq:isotropy}
\frac{\Delta r_{\parallel}}{\Delta r_{\bot}}={\rm true\ ratio}.
\end{equation}
This relation is violated when incorrect cosmological parameters are adopted.
The deviation of $\Delta r_{\parallel}/\Delta r_{\bot}$ from the true ratio enables us to quantify to what extent the adopted cosmology deviates from the correct one.
Since we only care about shape, the intrinsic size of object does not need to be known.

From Equation (\ref{eq:distance}) and Equation (\ref{eq:isotropy}) we find that AP test depends on the quantity
\begin{equation}
\label{eq:Fz}
F(z)\equiv\frac{(1+z)}{c}D_A(z)H(z).
\end{equation}
$\Delta z$ and $\Delta \theta$ are observed quantities, and do not change.
If we adopt a wrong cosmology to convert galaxy redshift $z$ to comoving distance $r$,
the ratio $\Delta r_\parallel/\Delta r_\bot$ will change and the degree
of distortion is given by
\begin{equation}
\label{eq:Fz}
\frac{[\Delta r_\parallel/\Delta r_\bot]_{\rm wrong}}
{[\Delta r_\parallel/\Delta r_\bot]_{\rm true}} =
\frac{[D_A(z)H(z)]_{\rm true}}{[D_A(z)H(z)]_{\rm wrong}}.
\end{equation}
For any object with known fixed ratio of the sizes along and across the LOS
one can use this relation to constrain the cosmological parameters
governing the expansion of the Universe.


We provide in Figure \ref{fig_AP} a demonstration of the deviations from isotropy for incorrectly chosen cosmological parameters.
We assume that the true cosmology corresponds to a $\Lambda$CDM with $\Omega_m=0.26$.
Now we measure objects using the redshift-distance relations in four different cosmologies:
\begin{itemize}
 \item ``SCDM'': Standard Cold Dark Matter cosmology with $\Omega_m=1.0$.
 \item ``de Sitter'': de Sitter Universe with $\Omega_m=0$, $w=-1$.
 \item ``Quintessence'': quintessence-like dark energy component, $\Omega_m=0.26$ and $w=-0.5$.
\item ``Phantom'': phantom-like dark energy component, $\Omega_m=0.26$ and $w=-3.0$.
\end{itemize}
The left panel of Figure \ref{fig_AP} shows $F(z)$ in these cosmologies
(normalized by its value in the correct cosmology $F_{\rm fid}(z)$).
Note that $F_{\rm fid}/F$ characterizes the magnitude of the distortion.
For instance, in the case of de Sitter cosmology $F_{\rm fid}/F\approx1.3$ at $z=1.0$,
meaning that there is a 30\% stretch of size in the radial direction relative to tangential direction.

In the right panel of Figure \ref{fig_AP} we show the apparent shape of two squares in the four cosmologies
as measured by an observer located at the origin.
The distortions in different cosmologies are clearly shown.
SCDM and Quintessence cosmologies results in $F(z) > F_{\rm fid}(z)$, i.e.,
apparent compression along the LOS.
The opposite trend is observed in the de Sitter cosmology with a stretch along the LOS.

More importantly, Figure \ref{fig_AP} highlights the {\it redshift dependence} of AP distortion.
In the de Sitter cosmology, the radial stretch becomes stronger with increasing redshift,
while in the SCDM cosmology the trend is opposite.
As will be discussed later, this fact is of essential importance for our method,
making anisotropies induced by AP distinguishable from those induced by RSD.

In this paper the AP test is applied to the density gradient field,
which should be statistically isotropic on all scales when the conversion from observed galaxy redshifts to comoving distances is correctly made.
Any anisotropy in the density gradient field and the variation of the degree of anisotropy with redshift are evidence for incorrectly adopted cosmology.
Sensitivity of the anisotropy to the cosmological parameters comes through the product $D_{\rm A}(z)H(z)$.

\section{The HR3 Mock Surveys}

We test our method using mock surveys constructed from the HR3.
The Horizon Runs are a suite of large volume N-body simulations with resolutions enough to a few major redshift surveys \citep{park 2005,horizonrun}.
HR3 adopts a flat-space $\Lambda$CDM cosmology with the WMAP 5 year parameters
$\Omega_{m}=0.26$, $H_{0}=72{\rm km/s/Mpc}$, $n_{s}=0.96$ and $\sigma_8=0.79$ \citep[]{komatsu 2011}.
The simulation was made in a cube of volume $(10.815 {~ h^{-1}} {\rm{Gpc}})^3$
using $7120^3$ particles with particle mass of $1.25\times 10^{11}$\hMsun.

The simulations started at $z=27$ and reached $z=0$ after making $N_{\rm step}=600$ timesteps.
Dark matter halos are identified using the Friend-of-Friend algorithm with the linking length of 0.2 times the mean particle separation.
Then the Physically Self Bound (PSB) subhalos that are gravitationally self-bound and tidally stable are identified \citep{kim and park 2006}.

An all-sky, very deep light cone survey reaching redshift $z=4.3$ was made by an observer located at the center of the box.
The co-moving positions and velocities of all CDM particles are saved as they cross the past light cone
and PSB subhalos are identified from this particle data.
To match the observations of recent LRG surveys \citep{choi 2010,gott 2009, gott 2008},
a volume-limited sample of halos with constant number density of $3 \times 10^{-4} (h^{-1}{\rm Mpc})^{-3}$
are selected with varied minimum halo mass limit along with redshift. 
The light cone survey sample consists of subhalos at different redshifts, 
and their redshift dependence on comoving distance and evolution of clustering are automatically included. 
The peculiar velocity of the most-bound particle in each subhalo is set to that of the subhalo.

We divide the whole-sky survey sample into four equal sky area subsamples 
and impose a maximal distance cut of 3000 Mpc/h.
This creates four $1/4$ sky surveys reaching $z=1.4$.
We further impose a minimal distance cut of $r>500$ Mpc/h (or equivalently $z>0.17$),
which is equal to that of the BOSS LOWZ sample. The BOSS LOWZ sample is usually restricted to
$z>0.15$ where the galaxy number density is more or less uniform \citep{Tojeiro2011,Tojeiro2012,Parejko2013}.

\section{The Density Gradient Field Distorted by AP and RSD}

For each mock survey, we embed the volume into a $250\times250\times500$ grid,
and estimate the density gradient vectors at each cell from
\begin{eqnarray}
 &\rho({\bf r}) = \sum_i m_i W({\bf r}-{\bf r}_i,h),\\
 &\nabla\rho({\bf r}) = \sum_i m_i \nabla W({\bf r}-{\bf r}_i,h),
\end{eqnarray}
where $\rho({\bf r})$ is the halo mass density at position ${\bf r}$, $m_i$ is the mass of the $i$-th halos,
and $W$ is the smoothing kernel,
for which we choose the 3rd order B-spline functions having non-zero value within a sphere of radius $2 h\ h^{-1}{\rm Mpc}$ \citep{GM1977,Lucy1977}.
We adopt a variable radius of the smoothing kernel so that the kernel
includes 20 nearest neighbor halos within $2h$.
In our sample of halos with the mean comoving number density of halos is $3\times 10^{-4} (h^{-1} {\rm Mpc})^{-3}$,
the typical value of $h$ is 12.5.
We find that the results of our method is rather insensitive to the choice of halo mass density or halo number density,
therefore in this paper we only present results based on the halo mass density field.


To quantify the anisotropy we use the angle between the density gradient vector and the LOS direction, $\theta$, where we define
\begin{equation}
 \mu\equiv |\cos \theta| = \frac{|{\bf r}\cdot\nabla\rho({\bf r})|}{|{\bf r}|\times|\nabla\rho({\bf r})|}.
\end{equation}
For an isotropic field with gradient vectors uniformly sampled within the survey volume,
$\mu$ follows a uniform distribution within [0,1].
To characterize the isotropy of the whole gradient field we look at the mean value of gradient vectors
\begin{equation}
 \bar\mu \equiv {\sum_{i=1,... n_{\rm vector}} \mu_i}/n_{\rm vector},
\end{equation}
where $n_{\rm vector}$ is the total number of gradient vectors.
An isotropic field has $\bar\mu=0.5$,
while a compression or stretch along LOS results in $\bar\mu>0.5$ or $\bar\mu<0.5$, respectively.

\begin{figure*}
   \centering{
   \includegraphics[width=8.1cm]{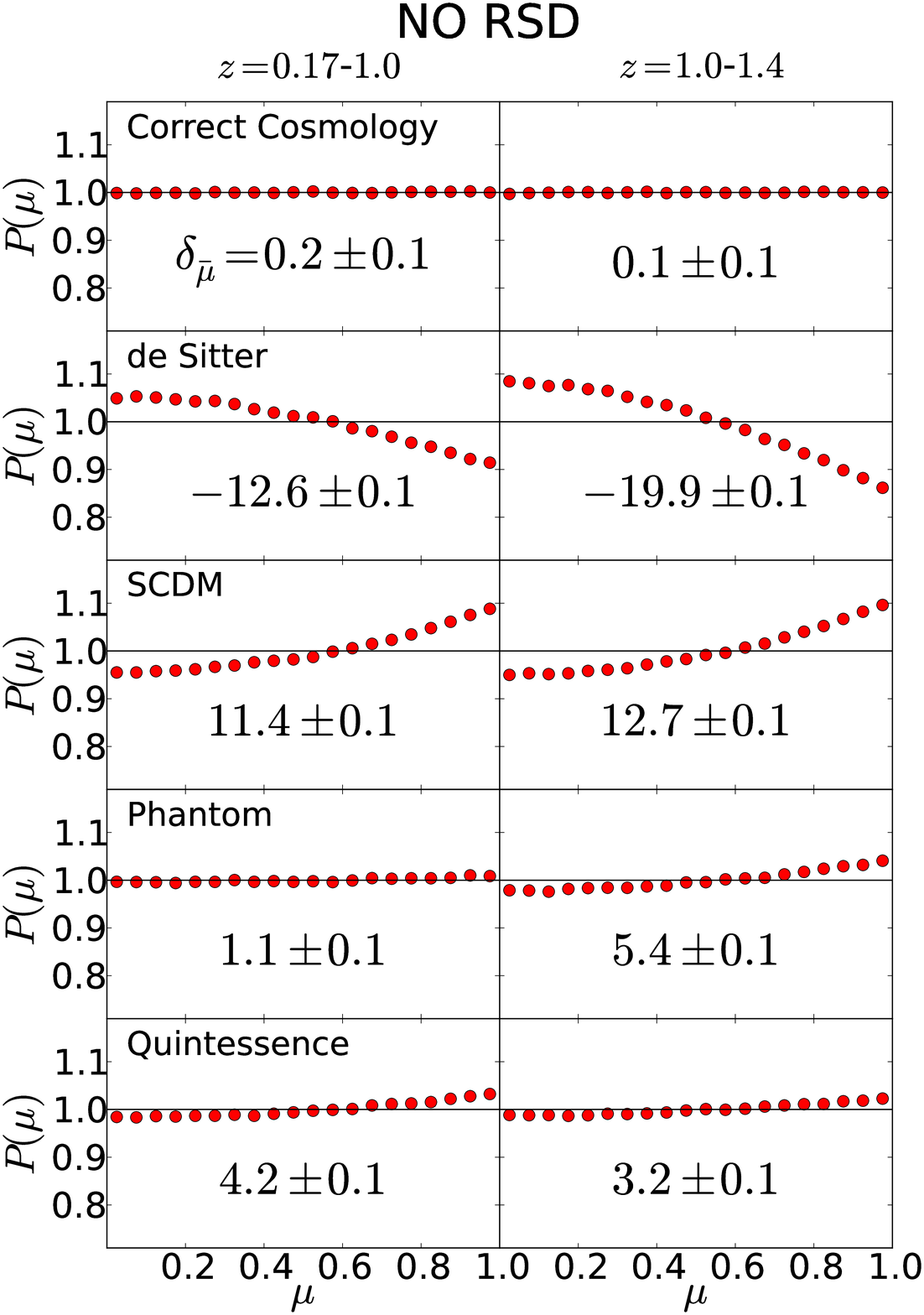}
   \includegraphics[width=8.1cm]{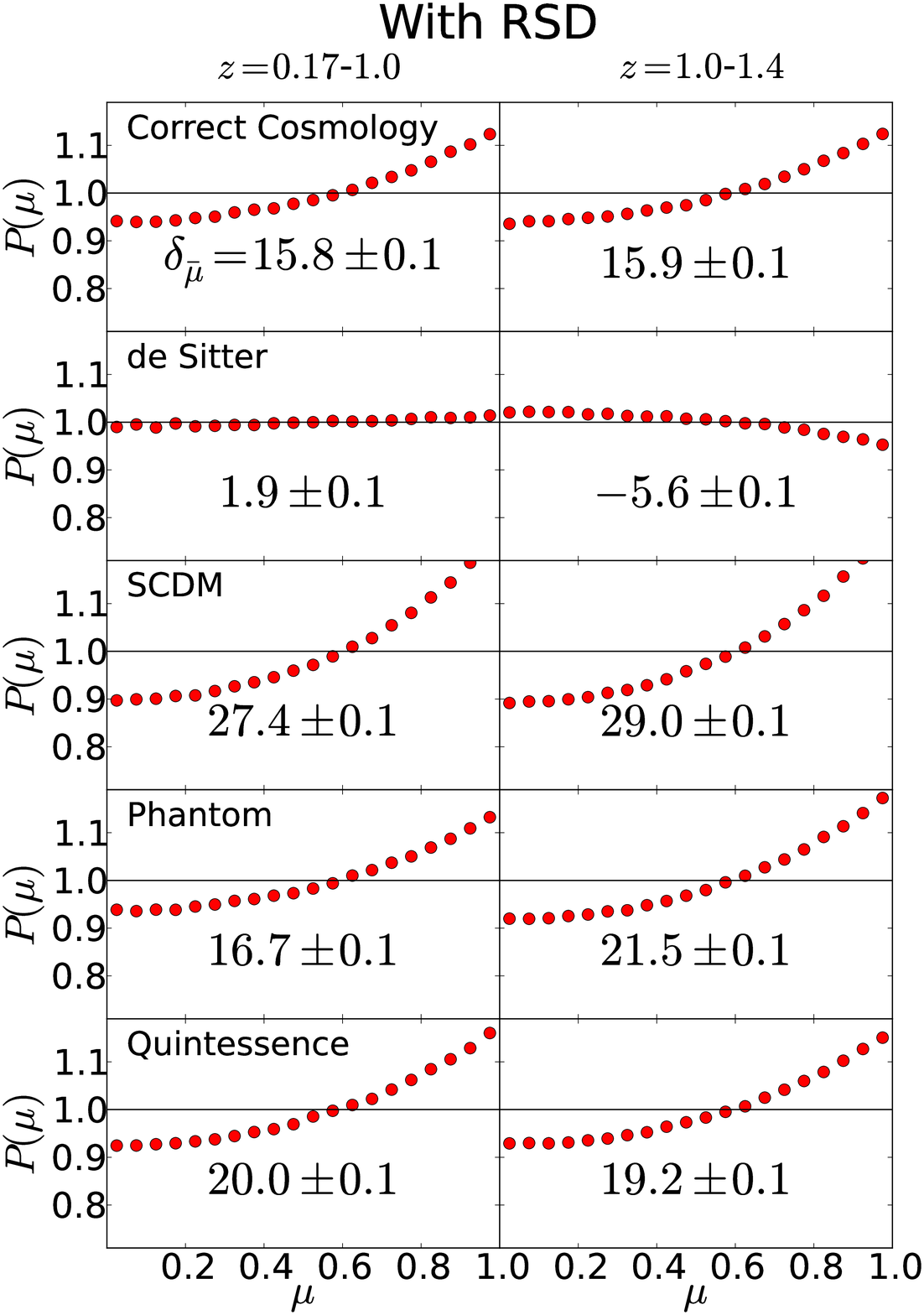}}
   \caption{\label{fig_muhists}
   Distributions of $\mu=\cos \theta$ measured from the density gradient fields of the HR3 mock surveys,
   assuming five different cosmologies.
   $\theta$ is the angle between the gradient vector and the line-of-sight.
   No RSD and with RSD results are shown in left and right panels, respectively.
   Gradient vectors lying within the redshift ranges of $0.17<z<1.0$ and $1.0<z<1.4$ are plotted separately to show the redshift dependence.
   To show the deviation of $\bar\mu$ from 0.5 we list the values of $\delta_{\mu}\equiv(\bar\mu-0.5)\times10^3$ in all panels.
   Anisotropies induced by the RSD effects are large but their redshift dependence is much smaller than those induced by the AP effect.
   We can find out the true cosmology by requiring that $\bar\mu$ has uniform values at both low and high redshifts.}
\end{figure*}

In Figure \ref{fig_muhists} we present histograms of $\mu$ measured from the mock surveys.
We adopt the correct cosmology and the four wrong cosmologies mentioned in section 2,
and compute values of $\mu$ in two different redshift ranges $z=0.17-1.0$ and $z=1.0-1.4$.
Left/right panel shows the results without/with the effect of RSD, respectively.

\subsection{AP effect without RSD}

%

In the left panel of Figure \ref{fig_muhists} we calculate values of $\mu$ without considering the RSD effect.
So if the obtained distribution is not uniform,
the only reason is that we are using a wrong cosmology to calculate galaxy distances.

As expected, we find the correct choice of cosmology leads to uniformly distributed $\mu$ with $\bar\mu\approx 0.5$,
while in wrong cosmologies this uniform distribution is not obtained.
In SCDM and Quintessence cosmologies, the apparent compression of structures along the LOS enhances the distribution at large $\mu$,
results in $\bar\mu>0.5$. 
Similarly, de Sitter and Phantom cosmologies give $\bar\mu<0.5$.

Comparing the low redshift and high redshift histograms we find redshift dependence of the anisotropy when incorrect cosmologies are adopted.
For instance, a choice of the de Sitter cosmology results in $\bar\mu= 0.4874$ and 0.4801 for $0.17<z<1.0$ and $1.0<z<1.4$, respectively.
Namely, the high-redshift region shows a larger deviation from 0.5, the isotropic case, as expected from Figure \ref{fig_AP}.
This results in a detection of redshift dependence at 48$\sigma$ CL.
Similarly, SCDM, Phantom and Quintessence cosmologies show redshift dependence at 8.1$\sigma$, 29$\sigma$ and 7.0$\sigma$ CLs.

\subsection{AP effect with RSD}

As the distances of galaxies are estimated from their redshifts in the actual situation,
there exists a systematic bias in the distribution of galaxies.
On small scales, high-density regions are stretched along the LOS due to the random motions of galaxies.
On large scales, the large-scale peculiar velocity field produces LOS compression of filaments and walls and radial elongation of voids.
On the smoothing scales we are interested in, the latter effect is more important.
To incorporate the RSD effects in the distribution of galaxies in our mock survey samples we change the radial distances of galaxies using the formula
\begin{equation}
r=\int_0^{z_{\rm cosmo}+\Delta z} \frac{cdz^\prime}{H(z^\prime)},\ \Delta z = \frac{v_{\rm LOS}}{c}(1+z_{\rm cosmo}),
\end{equation}
where $z_{\rm cosmo}$ is the cosmological redshift of the galaxy, and $v_{\rm LOS}$ is the LOS component of the proper galaxy peculiar velocity.
The distribution of $\mu$, after taking into account the RSD effects, is shown in the right panel of Figure \ref{fig_muhists}.

\begin{figure*}
   \centering{
   \includegraphics[height=6.0cm]{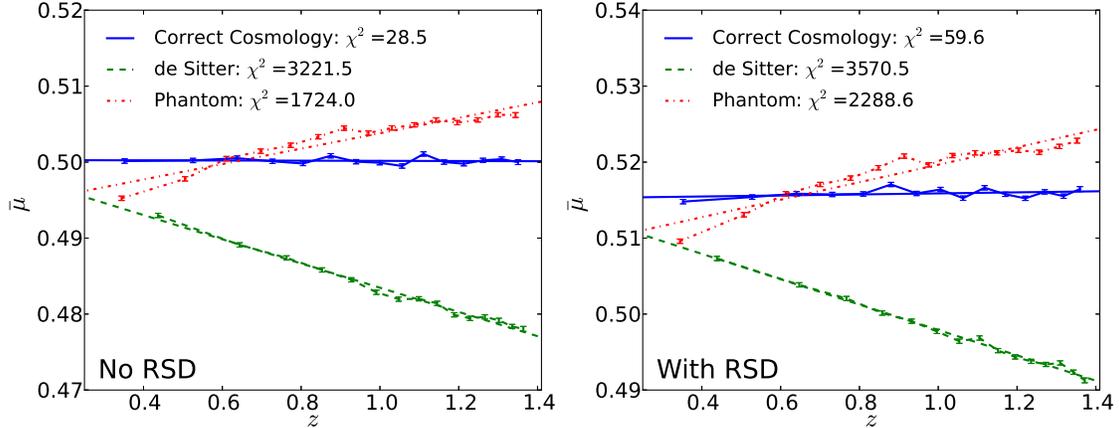}
   \caption{\label{fig_binnedmu}
   $\bar\mu$ as a function of redshift measured in a mock survey drawn from HR3.
   Three different cosmologies are adopted to compute the positions of galaxies.
   Results are shown for cases without and with RSD in left and right panels, respectively.
   The lines are linear fits to the points (note that the relations are not always linear).
   $\chi^2$ values are calculated from Equation (\ref{eq:chisq1}).
   }}
\end{figure*}

\begin{figure*}
\centering{
   \includegraphics[height=6.0cm]{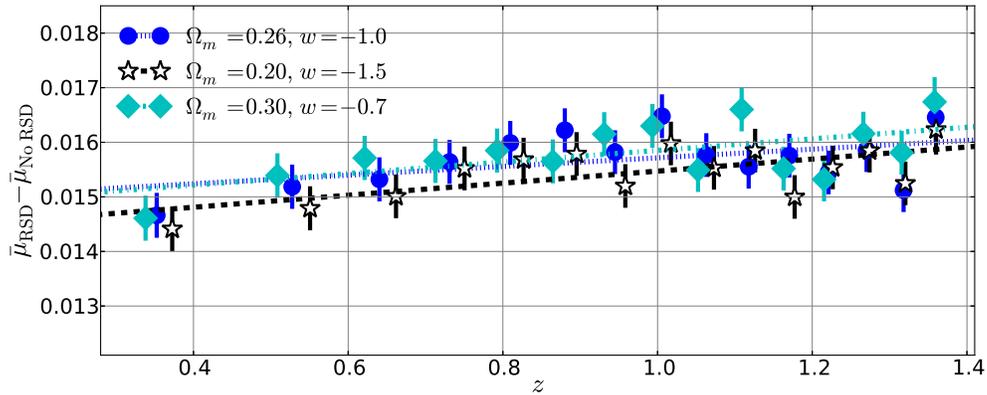}
   \caption{\label{fig_deltamu}
   Redshift dependence of the RSD effect,
   characterized by the quantity $\Delta \bar\mu \equiv \bar\mu_{\rm RSD}-\bar\mu_{\rm No\ RSD}$
   and measured in the correct (blue dots) and two incorrect cosmologies with parameters moderately deviated from the correct values.
   The lines are linear fits to the points.
   We find the slops of these lines are rather small, so RSD does not introduce significant redshift dependence into $\bar\mu$.
   Fitted lines are roughly parallel to each other,
   suggesting a small cosmological dependence of the redshift dependence of the RSD effects.}}
\end{figure*}

We find that the degree of the anisotropies produced by RSD is very large. 
In all cosmologies we find $\bar\mu>0.5$ with CL$>40$, which means that RSD overwhelms AP.
This makes us impossible to correctly determine the right cosmology by simply requiring $\bar\mu=0.5$ to the distribution.


On the other hand, we note that the redshift dependence of $\bar\mu$ is not significantly affected by RSD.
In the correct cosmology, the difference between the $\bar\mu$s of nearby and farther volumes is as small as $0.0001$,
on the same level of statistical fluctuation.
On the other hand, de Sitter, SCDM, Phantom and Quintessence cosmologies all show an evident redshift dependence of $\bar\mu$ at
50$\sigma$, 10.4$\sigma$, 33$\sigma$ and 5.5$\sigma$ CLs, respectively, which are close to the CLs of the corresponding results with no RSD effect.
The fact that {\it the effect of RSD is large but its redshift dependence is small} makes our method still applicable for the data with RSD.
Even with RSD, we can still correctly find out the true cosmology by using the relative change of the gradient field anisotropy with redshift.


\section{Likelihood of the Galaxy Density Gradient Field}

In the last section we showed that incorrect cosmologies result in redshift dependent $\bar\mu$,
a phenomenon less affected by RSD.
Inspired by this fact we construct the following likelihood function to discriminate between different cosmologies
\begin{equation}\label{eq:chisq1}
\chi^2\equiv \sum_{i=1}^{n_{\rm bin}} \left(\frac{\bar\mu_i-\bar\mu_{\rm whole}}{\sigma_{\bar\mu_i}}\right)^2.
\end{equation}
We split the sample into $n_{\rm bin}$ redshift bins having an equal comoving volume,
compute the values of $\bar\mu$ in each redshift bin,
and quantify to what extent they deviate from the $\bar\mu$ averaged over all redshift bins.
The value of $n_{\rm bin}$ shall be chosen according to the redshift range of the sample.

Figure \ref{fig_binnedmu} shows the $\chi^2$s calculated based on $15$ redshift bins measured in one of our mock surveys with redshift range from 0.17-1.4
\footnote{We keep the comoving volume of each bin the same. Therefore, the centers of the bins are not equally spaced,
and also depend on the adopted cosmology.}.
Three different cosmologies, the correct, de Sitter and Phantom cosmologies, are adopted.

The results without and with RSD effects are shown in left and right panels, respectively.
In both cases, we find the correct cosmology results in nearly a uniform value of $\bar\mu$ in different redshift bins,
while the de Sitter/Phantom cosmology has decreasing/increasing $\bar\mu$s with increasing redshift.
Wrong cosmologies are strongly disfavored with $\chi^2>1700$ while the correct cosmology has $\chi^2<60$.
Since the RSD effect enhances the values of $\bar\mu$ but does not significantly alter its redshift dependence,
the correct cosmology can be found regardless of RSD 
\footnote{If we restrict the redshift range to $z<0.8$ (roughly the maximum redshift of BOSS),
the de Sitter and Phantom cosmologies result in $\chi^2$s of 145.5/142.7 and 221.1/325.5 for cases without/with RSD effects, respectively.}.

\begin{figure*}
   \centering{
   \includegraphics[height=7.5cm]{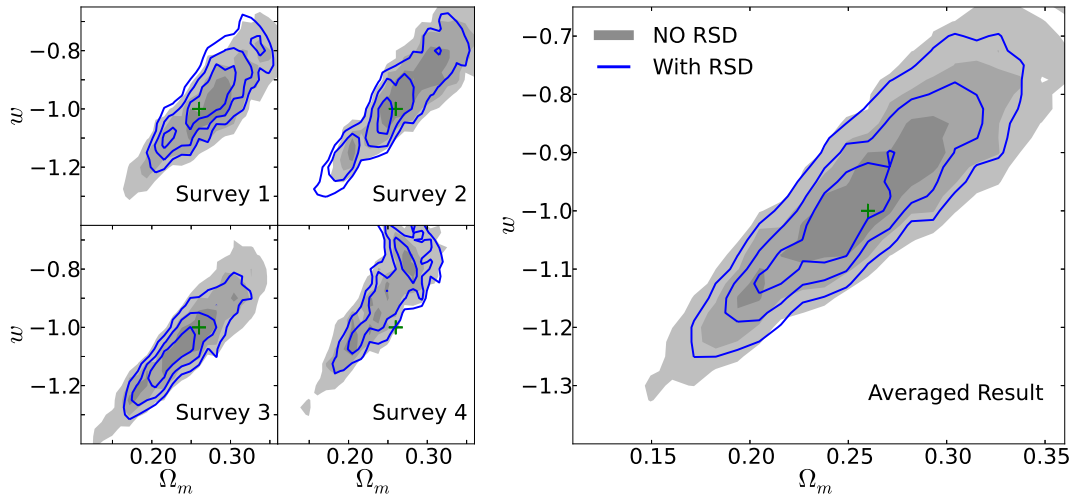}
   \caption{\label{fig_Contour}
   Likelihood contours (68.3\%, 95.4\% and 99.7\%) in the $\Omega_m$-$w$ plane,
   obtained from the $1/4$ sky HR3 mock surveys with a redshift range from 0.17 and 1.4.
   Left panel shows the results of four individual mock surveys.
   Right panel shows the average.
   No RSD and with RSD contours are plotted as gray filled regions and blue lines, respectively.
   The green cross marks the true cosmology.
   We achieve unbiased estimations of $\Omega_m$ and $w$ regardless of RSD.
   }}
\end{figure*}

The effect of RSD on $\bar\mu$ can be characterized by the quantity
\begin{equation}\label{eq:deltamu}
 \Delta \bar\mu \equiv \bar\mu_{\rm RSD}-\bar\mu_{\rm No\ RSD}.
\end{equation}
Figure \ref{fig_deltamu} shows $\Delta\bar\mu$ computed in three cosmologies,
i.e. the correct cosmology and two other cosmologies whose parameters moderately deviate from correct values.
Fitting $\Delta\bar\mu(z)$ to a linear function we find the redshift dependence of $\Delta\bar\mu$ is very small in all cosmologies.
Although the amplitudes of $\Delta\bar\mu$ in the three cosmologies are slightly different,
the effects of RSD in the three different cosmologies can be effectively removed 
since we are concerned with only the redshift dependence of $\bar\mu$.

To remove the remaining weak redshift dependence of RSD effects completely we modify the $\chi^2$ function as follows
\begin{equation}\label{eq:chisq2}
\chi^2\equiv \sum_{i=1}^{n_{\rm bin}} \left[\frac{\bar\mu_i-\bar\mu_{\rm whole} - (\Delta\bar\mu_i-\Delta\bar\mu_{\rm whole}) }{\sigma_{\bar\mu_i}}\right]^2.
\end{equation}
As an approximation, we use $\Delta \bar\mu$ computed in the correct cosmology in Equation (\ref{eq:deltamu}).
Figure \ref{fig_deltamu} demonstrates that the redshift dependence of $\Delta \bar\mu$ is insensitive to the change in cosmological parameters.
In principle $\Delta \bar\mu$ can be numerically estimated accurately for any trial cosmology.

Figure \ref{fig_Contour} shows the likelihood of cosmological models in the $\Omega_m$ and $w$
space obtained by computing Equation (\ref{eq:chisq2}) in the four individual mock surveys (panels on the left) and their average (right panel).
We find that the correct estimation of $\Omega_m$ and $w$ is achieved in both cases with and without RSD.
The constrained regions of the two cases are close to each other,
meaning that RSD contamination is completely removed.

We find $\Omega_m$ and $w$ are positively degenerated with each other. This is expected.
For instance, reducing $\Omega_m$ and having a more phantom-like dark energy produce similar influences on the expansion history of the Universe.
Roughly, we find $\Omega_m$ and $w$ are constrained to $0.25\pm0.05$ and $-1.0\pm0.1$ (68.3\% CL), respectively,
using the sample like ours and our AP test method alone.


\section{Discussion and Conclusions}

In this paper we propose a novel method to use the Alcock-Paczynski test applied to the galaxy density gradient field.
If an incorrect cosmology is chosen to compute the distances of galaxies from redshifts the gradient field appears to be anisotropic with the degree of anisotropy varying with redshift.
RSD effects also produce large anisotropy in the gradient field but maintain a roughly uniform magnitude at all redshifts.
By focusing on the redshift dependence of degree of anisotropy,
we are able to derive correct estimations of cosmological parameters in spite of contamination induced by RSD.

Our method is a new attempt to apply the AP test to the large scale structure of the Universe. 
It is complementary to the existing AP tests using the 2-point correlation function, galaxy pairs, 
and large scale voids, or to the methods for measuring the cosmic expansion history, 
e.g., type Ia supernovae, baryon acoustic oscillations, and topology \citep{topology} in three ways. 
First, it uses the galaxy density gradient field, a new approach to apply the AP test. 
Second, since our method is using the redshift dependence of $\bar\mu$, and not its absolute magnitude, 
we are measuring the first derivative of $D_AH$, while other methods mainly focus on $D_AH$. 
Our method utilizing the redshift dependence of the AP effect can be combined with other methods to 
take full advantage of the cosmological information encoded in the large scale structure data. 
It should be also pointed out that our method allows us to use a given observational data down to scales of about $20h^{-1}$Mpc, 
much smaller than that of currently popular BAO method (about 100$h^{-1}$Mpc). 
Third, Figure \ref{fig_binnedmu} and \ref{fig_deltamu} show that the redshift dependence of $\bar\mu$ has only a very weak dependence on RSD, 
which can be effectively removed by using mock data. 
The change of $\bar\mu$ with redshift is dominated by the systematic effects of the assumed cosmology, 
and the cosmological parameter estimation does not suffer from the bias due to the RSD.
 
One might worry about the systematic effects due to the galaxy sample variation with redshift. 
A sample of galaxies whose bias changes with redshift can have the LOS density gradient somewhat affected by the sampling variation. 
However, the selection effects of the density tracers will not much affect the results of our method because of the following reasons. 
(1) Our method uses only the ``local isotropy'' of the density field of the tracer. 
Unless the target selection varies significantly over the scale of smoothing ($<$50Mpc/h), 
our method will not be affected by the variation of bias in the galaxy distribution. 
(2) As can be seen in Figure \ref{fig_binnedmu} using our mocks, 
the systematic change in the selection of the tracer galaxies does not much affect the shape of $\bar\mu(z)$ 
when a volume-limited (constant comoving number density) sample with the minimum mass cut varying with redshift is used. 
The shape of the blue line on the right panel (with RSD) is almost the same as that in the left panel (no RSD) 
even though the mock galaxies at $z\sim0.17$ have much larger mass ($\sim1.4\times10^{14} M_{\sun}$) than those at $z\sim1.4$
($\sim6\times10^{13} M_{\sun}$). 
The difference is in Figure \ref{fig_deltamu} 
(The variation of the minimum halo mass giving a constant number density of BOSS, 
for example, can be found in Figure 6 of \cite{horizonrun}). 
The small residual RSD effects on the density field can be different for different bias galaxies, 
and this is numerically estimated and subtracted by using the mock samples.
 
When dealing with real observational data, 
the sampling bias can vary more widely than in our mock sample and 
it will be important to accurately model the observed galaxies to remove the small residual RSD effects on the isotropy of smoothed density field. 
It is also needed to handle various observation-related effects such as survey geometry, selection bias, fiber collisions, etc. 
We will report results of such investigations in forthcoming studies.

\section*{Acknowledgments}

We thank the Korea Institute for Advanced Study for providing computing resources
(KIAS Center for Advanced Computation Linux Cluster System).
We thank Seokcheon Lee, Cristiano Sabiu and Hyunmi Song for helpful discussions.
JEFR acknowledges financial support from Vicerrectoria de Investigaciones through a FAPA project.

\end{document}